\documentclass[12pt,preprint]{aastex}

\newcommand{\myemail}{courteau@astro.ubc.ca}
\newcommand{\etal}{et~al.~}

\newcommand{\hub}{H$_{\hbox{\scriptsize 0}}$}
\newcommand{\kms}{\ifmmode\,{\rm km}\,{\rm s}^{-1}\else km$\,$s$^{-1}$\fi}
\newcommand{\magarc}{\ifmmode {{{{\rm mag}~{\rm arcsec}}^{-2}}}
             \else {{{mag}$~${arcsec}$^{-2}$}}
             \fi}

\def \spose#1{\hbox to 0pt{#1\hss}}
\def \lta{\mathrel{\spose{\lower 3pt\hbox{$\sim$}}
     \raise 2.0pt\hbox{$<$}}}
\def \gta{\mathrel{\spose{\lower 3pt\hbox{$\sim$}}
     \raise 2.0pt\hbox{$>$}}}

\def\Equ#1{Eq.~(\ref{eq:#1})}
\def\se#1{\S\ref{sec:#1}}
\def\Fig#1{Fig.~\ref{fig:#1}}

\def\eg{{e.g.~}}
\def\ie{{i.e.~}}

\def\be{\begin{equation}}
\def\ee{\end{equation}}

\def\ifm#1{\relax\ifmmode#1\else$\mathsurround=0pt #1$\fi}
\def\kms{\ifmmode\,{\rm km}\,{\rm s}^{-1}\else km$\,$s$^{-1}$\fi}

\def\ltsima{$\; \buildrel < \over \sim \;$}
\def\lsim{\lower.5ex\hbox{\ltsima}}
\def\gtsima{$\; \buildrel > \over \sim \;$}
\def\gsim{\lower.5ex\hbox{\gtsima}}

\def\pmb#1{\setbox0=\hbox{#1}%
\kern-.025em\copy0\kern-\wd0
\kern.05em\copy0\kern-\wd0
\kern-.025em\raise.0433em\box0}

\def \ion#1#2{#1{\footnotesize{#2}}\relax}
\def \ha       {H$\alpha$\ }
\def \hi       {\ion{H}{I}\ }

\def \ltsima{$\; \buildrel < \over \sim \;$}
\def \simlt    {\lower.5ex\hbox{\ltsima}}
\def \farcm{\hbox{$.\mkern-4mu^\prime$}}
\def \farcs{\hbox{$.\!\!^{\prime\prime}$}}
\def \Rx        {R_{exp}}
\def \V22{V_{disk}}
\def \r22{R_{disk}}

\def \VI        {V-I}

\def \dlogV     {\partial\log{\V22}}
\def \dlogR     {\partial\log{\Rx}}
\def \partialvr {\dlogV \thinspace / \thinspace\dlogR}
\def \littlemm{\ifmmode{\scriptscriptstyle m }
     \else{\hbox{$\scriptscriptstyle m $ }}\fi}
\def \topemm{\raise .9ex \hbox{\littlemm}}
\def \magpoint{\hbox to 2pt{}\rlap{\hskip -.5ex \topemm}.\hbox to 2pt{}}
\def \deg {$^\circ$}

\shorttitle{The Tully-Fisher Relation of Barred Galaxies}
\shortauthors{Courteau, Andersen, Bershady, MacArthur, \& Rix}
\slugcomment{For publication in ApJ (v594).}
\begin{document}

\title{The Tully-Fisher Relation of Barred Galaxies}

\author{St\'{e}phane Courteau\altaffilmark{1}}

\affil{Department of Physics \& Astronomy, University of British Columbia,
    6224 Agricultural Road, Vancouver, BC V6T 1Z1}
\email{\myemail}

\author{David R. Andersen\altaffilmark{1}}
\affil{Max-Planck-Institut f\"{u}r Astronomie, K\"{o}nigstuhl 17,
     D-69117 Heidelberg, Germany}
\email{andersen@mpia-hd.mpg.de}

\author{Matthew A. Bershady}

\affil{Univ. of Wisconsin, Dept. of Astronomy, 475 N. Charter St.,
       Madison, WI 53706}
\email{mab@astro.wisc.edu}

\author{Lauren A. MacArthur\altaffilmark{1}}
\affil{Department of Physics \& Astronomy, University of British Columbia,
    6224 Agricultural Road, Vancouver, BC V6T 1Z1}
\email{lauren@astro.ubc.ca}

\and

\author{Hans-Walter Rix}
\affil{Max-Planck-Institut f\"{u}r Astronomie, K\"{o}nigstuhl 17,
     D-69117 Heidelberg, Germany}
\email{rix@mpia-hd.mpg.de}

\altaffiltext{1}{
Visiting Astronomer, Kitt Peak National Observatory, National Optical 
Astronomy Observatory, which is operated by the Association of 
Universities for Research in Astronomy, Inc. (AURA) under cooperative 
agreement with the National Science Foundation. }

\altaffiltext{1}{    }

\begin{abstract}
We present new data exploring the scaling relations, such as the
Tully-Fisher relation (TFR), of bright barred and unbarred galaxies.
A primary motivation for this study is to establish whether barredness
correlates with, and is a consequence of, virial properties of galaxies. 
Various lines of evidence suggest that dark matter is dominant in 
disks of bright {\it unbarred} galaxies at 2.2 disk scale lengths, 
the point of peak rotation for a pure exponential disk. We 
test the hypothesis that the TF plane of {\it barred} high surface 
brightness galaxies is offset from the mean TFR of unbarred galaxies, 
as might be expected if barred galaxies are ``maximal'' in their 
inner parts.  We use existing and new TF data to search for basic 
structural differences between barred and unbarred galaxies. 
Our new data consist of 2-dimensional H$\alpha$ velocity 
fields derived from SparsePak integral field spectroscopy (IFS) and 
V,I-band CCD images collected at the WIYN Observatory\footnote{The 
WIYN Observatory is a joint facility of the University of 
Wisconsin-Madison, Indiana University and the National Optical
Astronomy Observatory.} for 14 strongly
barred galaxies. Differences may exist between kinematic and photometric 
inclination angles of barred versus unbarred galaxies.
These findings lead us to restrict our analysis to barred galaxies 
with $i>50$\deg. We use WIYN/SparsePak (2-D) velocity fields to show that 
long-slit (1-D) spectra yield reliable circular speed measurements
at or beyond 2.2 disk scale lengths, far from any influence of the bar. 
This enables us to consider line width measurements from extensive
Tully-Fisher surveys which include barred and nonbarred disks and 
derive detailed scaling relation comparisons. 

We find that for a given luminosity, barred and unbarred galaxies have 
comparable structural and dynamical parameters, such as peak velocities, 
scale lengths, or colors.  In particular, the location of a galaxy in the 
TF plane is independent of barredness.  In a global dynamical sense, 
barred and unbarred galaxies behave similarly and are likely to have, 
on average, comparable fractions of luminous and dark matter at a given 
radius.  

\end{abstract}


\section{Introduction}\label{sec:intro}

Based on the flatness of rotation curves in spiral galaxies and the
density profiles inferred from X-ray temperatures and stellar velocity 
dispersion profiles in ellipticals, 
it is widely believed 
that galaxies are embedded in non-dissipative massive dark halos.  
More than 90\% of the total mass of a galaxy would be in the form 
of dark matter.  Less appreciated is the fact that we still have 
a rather muddled picture of the mass distribution of luminous 
and dark matter in the luminous part of a galaxy.  
This is unfortunate since the final distribution of baryons in
a galaxy is a tell-tale sign of its formation and evolution.  
Numerical and analytical models of disk formation in 
a dissipationless dark matter halo predict, for realistic total 
fractions of baryonic to dark matter, that spiral disks should 
live in dark halos that dominate the mass fraction at nearly 
all radii (\eg Mo, Mao, \& White 1998), beyond about a disk scale 
length.  This ratio may quite possibly be different for barred 
and unbarred galaxies of a given total mass or luminosity 
(Courteau \& Rix 1999; hereafter CR99). 

Recent debates about the Cold Dark Matter paradigm (\eg 
Weinberg \& Katz 2002; Sellwood 2003; Courteau \etal 2003a) and 
galaxy structural properties inferred from new infrared surveys 
(\eg Eskridge et~al 2002;  MacArthur \etal 2003) have brought 
barred galaxies to the fore.  Bar perturbations in 
galaxies, far from just being dynamical curiosities, actually play 
a fundamental role in shaping galaxies into the structures we see 
today (see Buta, Crocker, \& Elmegreen 1996 for reviews).  For instance,
the early dynamical evolution of a massive rapidly rotating gaseous bar
could provide enough energy and angular momentum to significantly modify
the inner CDM halo (Silk 2002).  Dynamical 
and structural studies of barred galaxies are however few, due in 
part to the complexity in interpreting their velocity fields 
(\eg Weiner \etal 2001) and their surface brightness profiles 
(\eg MacArthur \etal 2003).  Many large-scale flow studies of 
spiral galaxies have also excluded disturbed or barred galaxies
to minimize scatter, as previously believed, in the 
distance-measuring technique.
The latter studies have enabled extensive scaling relation studies of 
unbarred galaxies, but little attention has been paid to their barred 
cousins. This is again deplorable as a comparative study of the scaling 
relations for barred and unbarred galaxies would potentially 
unravel clues about the structure and origin of bars and 
the role of dynamical processes in establishing the Hubble sequence
of disk galaxies. 

The body of numerical simulations of barred galaxies is comparatively
richer and has recently reached new heights with the availability 
of superior N-body realizations with more than $10^6$ particles 
(post 2010 readers may enjoy a moment of laughter). 
Until just recently, it was believed that bar instabilities in 
a disk might be suppressed by a massive halo.  
Thus only low concentration halos, or equivalently systems of very 
high surface brightness (HSB) or low angular momentum per unit luminosity, 
would be prone to generating a non-axisymmetric (bar/oval) structure 
in their center (Ostriker \& Peebles 1973; accounts of the misconceptions
surrounding this argument are presented in Bosma 1996 and Sellwood \& Evans 
2001).

The suggestion that barred galaxies would have an especially high 
ratio of baryons-to-dark matter within the optical disk (``maximal 
disk'') might imply that these systems define their own sequence 
in the luminosity-line width diagram, if one assumes that unbarred 
galaxies are, on average, sub-maximal (CR99).  Thus, for a given 
absolute magnitude, a galaxy with higher baryon fraction, or disk 
mass-to-light ($M/L$) ratio, would have a shorter disk scale length
and rotate faster.  Verification of this important, though 
tentative, suggestion should be easily obtained from a large sample
of uniformly selected barred galaxies that are part of a 
well-calibrated, self-consistent luminosity-line width survey.  The 
current study was largely motivated by this question. 

In discussing the mass distribution in spiral galaxies, we shall 
use the definition that a disk is ``maximal'' if it contributes 
more than 75\% of the total rotational support of the galaxy at 
$\r22\equiv 2.2h_{disk}$, the radius of maximum disk circular 
speed (Sackett 1997). Thus, for a maximal disk, 
${\V22}/V_{total} \gta 0.75$, where
$V_{total}$ is the total amplitude of the rotation curve at $\r22$
and $\V22=V(\r22)$. 
Note that for ${\V22}/V_{total} = 0.7$ the disk and halo contribute
equally to the potential at $\r22$.  Large bulges for late-type galaxies 
make little difference for the computation of this quantity at $\r22$ (CR99). 

The pattern speeds of bars have been considered as a potential indicator 
of the relative fraction of dark matter in galaxy disks.  N-body 
simulations of bar formation in 
stellar disks suggest that dynamical friction from a dense dark matter
halo dramatically slows the rotation rate of bars in a few orbital periods
(Debattista \& Sellwood 1998, 2000; hereafter DS00).  Because bars are 
observed to rotate quickly, DS00 proposed that dark matter halos in HSB
galaxies must have a low central density; thus, their disks ought to be 
maximal.  These simulations were revisited by Valenzuela \& Klypin (2003; 
hereafter VK03) with similar N-body simulations (no gas) but with an order 
of magnitude improvement in the force resolution.  VK03 found that dynamical 
friction from transfer of angular momentum of the bar to the halo does
play a role but, contrary to DS00, that effect appears to be small. 
In addition, VK03 find that bars can form even in the presence of strong 
halos, and that stellar disks make a negligible contribution to the inner 
rotation curve (at $\r22$).  The bars modeled in DS00 also span nearly the 
entire disk whereas the observed bar-to-disk scale length ratio seldom 
exceeds 1.5, as also pointed out 
by VK03.  These authors find that mass and force resolution are critical 
for modeling the dynamics of bars, and the contentious results from DS00
would stem primarily from numerical resolution effects. However, the 
higher force resolution of VK03 induces numerical viscosity which may 
bring their results into question (J. Sellwood 2003; priv. comm.)!  
Free from the vagaries of numerical simulations, Athanassoula (2003)
uses analytical calculations to warn against the use of bar slowdown 
rate to set limits on the baryonic to dark matter fraction within the 
optical radius\footnote{Athanassoula (2003) finds that the bar slowdown 
rate depends not only on the relative halo mass at a given radius, but
also on the velocity dispersion of both the bulge and disk components.},
in agreement with Sellwood (2003).  This point is however moot since 
one cannot observe bar slowdown from a single snapshot.  
Athanassoula (2003) further suggests that the ratio of corotation to 
bar length may not be an adequate estimator of halo fraction, but the
model ratios (see her Figs. 12 \& 14) upon which these conclusions are 
drawn are inconsistent with observations. 

A complete picture of bar dynamics awaits a self-consistent 
treatment of both the stars {\it and} gas embedded in a cosmologically 
motivated halo.  These simulations should 
include dynamical friction and ultimately reproduce the fraction 
of strong bars detected in the infrared and predict the rate of bar 
slowdown and dissolution as a function of bulge/total brightnesses, 
time, and environment.  

The model-independent quest of the relative matter distribution in 
barred and unbarred galaxies is by no means straightforward either,
but is most significant as it provides a necessary constraint for 
the shape and amplitude of the dark matter density profile in the 
luminous part of a galaxy. Whether disks are maximal or not at 
$\r22$, the inner 1-2~kpc may 
be dominated by baryons in most galactic systems, including 
early and late-type HSB barred and unbarred spirals 
(\eg Broeils \& Courteau 1997; Corsini \etal 1999),
low surface brightness (LSB) galaxies 
(Swaters 1999; Swaters, Madore, \& Trewhella 2000; Fuchs 2002) and 
ellipticals (\eg Brighenti \& Mathews 1997; see also Ciotti 2000). 
Maximally massive disks in LSB galaxies may however require 
unrealistically high disk $M/L$ ratios (Swaters \etal 2000; Fuchs
2002), based on stellar population synthesis models. 

Also troublesome is our lack of knowledge about the distribution
of matter in our own Milky Way.
Whether it has a maximal disk (Gerhard 2002) or not (Dehnen \& Binney 1998;
Klypin, Zhao, \& Somerville 2002) is still a matter of debate. Crucial
elements for local mass density estimates include the precise contribution of
the massive central bar (\eg Zhao, Rich, \& Spergel 1996) or elongated bulge
(Kuijken 1995), an accurate measure of the disk scale length, and constraints
from microlensing towards the bulge. 

The determination of the relative fraction of visible and dark matter
in external barred and unbarred galaxies relies on our ability to 
determine stellar $M/L$ ratios accurately.  
The modeling of disk dynamical mass in barred galaxies relies heavily on 
the interpretation of the non-axisymmetric motions of ionized gas around 
the bar within the context of a hydrodynamical model.  This model does 
have a local potential, and hence the bar and disk $M/L$ are parameters 
of the model.  It is certainly a more complicated approach than using 
collisionless particles as dynamical tracers, as with stellar velocity 
dispersions, but the latter has its own complications as well (\eg 
Swaters \etal 2003). Significant improvements in mass modeling techniques 
for individual galaxies are expected with the development of stellar 
population synthesis models (Bell \& de Jong 2001) and dynamical 
constraints (Weiner, Sellwood, \& Williams 2001) to yield realistic 
$M/L$ ratios, and further constraints from cosmological simulations 
of dark halos to curtail disk-halo 
degeneracies (Dutton, Courteau, \& de Jong 2003). 

Various lines of circumstantial evidence for external systems favor 
dark matter halos that dominate the mass budget within 
$\r22$.   Arguments based on the stellar kinematics of galactic disks
(Bottema 1997), gas kinematics (Kranz, Slyz, \& Rix 2003), the stability 
of disks (Fuchs 2001) and the lack of correlated scatter in the 
Tully-Fisher relation (hereafter TFR; Tully \& Fisher 1977)
of unbarred LSB and HSB galaxies (CR99) suggest that, {\it on average}, disks 
with $V_{max}<200$ \kms\ are sub-maximal.  The two very different 
analyses by Bottema and CR99 both yield 
$\V22/V_{total}=0.6\pm 0.1$, or $M_{dark}/M_{total}=0.6 \pm 0.1$ 
for HSB galaxies at $\r22$.  
The geometry of gravitational lens systems, coupled with rotation curve 
measurements, can also be used to decompose the mass distribution of a 
lensing galaxy. This promising technique, pioneered by Maller \etal 
(2000), has been applied to the galaxy-lens system 2237+0305 by
Trott \& Webster (2002) who find $V_{disk}/V_{total}=0.57 \pm 0.03$, 
in excellent agreement with the studies above and predictions from
analytical models of galaxy formation (e.g. Dalcanton \etal 1997; 
Mo, Mao, \& White 1998).

While a consistent picture of galaxy structure is emerging in which a
dark halo dominates with
$M_{dark}/M_{total} \geq 0.6$ well into the optical disk,
a number of pro-maximal disk arguments are still found in the 
literature, citing evidence from the shapes and extent of rotation 
curves and mass modeling (see, e.g., Bosma 2002).
The match between {\it pure disk}
mass models and \ha rotation curves (\eg Broeils \& Courteau 1997; 
Seljak 2002; Jimenez \etal 2003) is usually satisfactory for spiral 
galaxies of different surface brightnesses and morphologies 
and has often been invoked as evidence for a maximal 
concentration of baryons relative to the dark matter inside the optical 
disk (Buchhorn 1992; Palunas \& Williams 2000). 
However, mass modeling with \ha rotation curves alone is not a uniquely
determined problem.  The equivalence of, or degeneracy between, the 
two descriptions -- pure disk versus sub-maximal disk + dark halo -- 
was demonstrated in Broeils \& Courteau (1997) and CR99 for a sample 
of 300 disk galaxies; residuals for the maximal or sub-maximal 
fits are indistinguishable.  Without an accurate estimate of
${M/L}_{disk}$, or external constraints on $V_{disk}/V_{virial}$ at $\r22$,
mass modeling cannot dissentangle maximal and sub-maximal disk models.

Our study of the dynamical structure of barred and unbarred galaxies 
will offer new insights in the debate of the maximal disk hypothesis 
in barred and unbarred galaxies.  However, we plan to revisit this 
controversial issue in a future presentation (Courteau \etal 2003a).  
Here we pursue our comparison of barred and unbarred galaxies in 
the context of global scaling relations. 

\subsection{Available galaxy samples}\label{sec:avail}

The study of scaling relations of barred galaxies, and tracing their
location in the TFR, requires that we utilize ``fundamental plane'' 
surveys of an ensemble of galaxies.  The ``Shellflow'' and ``SCII'' 
all-sky Tully-Fisher surveys of Courteau \etal\@(2000) and 
Dale \etal\@(1999) are useful in that respect. 
These surveys were designed to map the convergence of the
velocity field on $\sim 60h^{-1}$ Mpc scales while minimizing  
calibration errors between different telescopes in different hemispheres;
state-of-the-art TF calibrations are thus available in both cases.
Both surveys 
include line width and luminosity measurements for a small fraction 
of barred galaxies that can be used to study structural trends, 
provided the presence of bar does not bias these measurements. 
More details about the surveys will be given in \se{TFbar}. 

In order to calibrate existing long-slit spectra of barred galaxies
and initiate a comprehensive study of barred galaxy velocity fields, 
we have collected new deep V and I-band images and integral field 
\ha\ velocity fields of 14 strongly barred galaxies at the WIYN 3.5-m 
telescope.  We present the new data and velocity field analysis 
in \se{WIYN} and discuss possible limitations of the data, such as 
due to inclination uncertainties and non-circular motions.  
We then examine the location of barred and unbarred galaxies in 
the TF samples discussed above in \se{TFbar}.  We find that barredness 
does not play a role in the luminosity-line width and luminosity-size 
planes of spiral galaxies.  
In \se{conclusion}, we discuss future programs that may benefit the 
study of scaling relations in barred and unbarred galaxies. 

\section{A New WIYN Survey of Barred Galaxies}\label{sec:WIYN}

\subsection{Observations}\label{sec:data}

In March 2002, we obtained 2-D \ha velocity maps and deep V and 
I-band photometry at the WIYN 3.5-m (3 nights) and WIYN 0.9-m (2 nights), 
respectively for 14 strongly barred bright galaxies (SBb-SBc; 
$m_B \simlt 15$; see Table 1) and one unbarred spiral galaxy (NGC 3029).
The galaxies were selected according to the same criteria as the 
TF Shellflow survey of spiral galaxies, save 
the emphasis on the bar-like morphology.
Ultimately, we aim to calibrate our new data on the same system 
as Shellflow, a survey deficient in barred galaxies, 
to enable direct comparisons between barred and unbarred systems. 

Integral field spectroscopy (IFS), which is lacking in Shellflow and 
SCII, is required to fully characterize the velocity amplitudes of
the bulge, bar, and underlying disk, especially if non-circular 
velocities are conspicuous. We have obtained 2-D velocity 
maps with the SparsePak integral field unit (Bershady \etal 2003a). 
The SparsePak IFU is a fiber-optic array of 82 fibers mounted at the
Nasmyth f/6.3 focus imaging port on the WIYN 3.5-m telescope. SparsePak 
has 75 fibers arranged in a sparsely filled grid sub-tending an area
of 72\arcsec $\times$ 73\arcsec.  Each fiber has an
active core diameter of 4\farcs69 (500 $\mu$m); cladding and buffer
increase the total fiber diameter to 5\farcs6.  The filling factor
for the grid is $\sim 25$\% on average, but rises to $\sim 55$\% in the
inner 16\arcsec\ where the fibers are more densely packed. In addition
to the 75 fibers arranged in a square, another 7 fibers are spaced
around the square roughly $70\arcsec$--$90\arcsec$ from the center and
are used to measure the ``sky'' flux.  An example of the SparsePak
footprint is shown in \Fig{fibers}.

SparsePak feeds the WIYN Bench Spectrograph, a fiber--fed spectrograph 
designed to provide low to medium resolution spectra. 
We used the Bench Spectrograph camera (BSC) and 316 lines mm$^{-1}$
echelle grating in order 8 to cover
6500A $ < \lambda <$ 6900A, with a dispersion of 0.2A~pix$^{-1}$
(8.8 {\kms}~pix$^{-1}$) and an instrumental FWHM of 0.6A~(26.5~\kms).
The BSC images the spectrograph onto a T2KC thinned SITe $2048 \times 2048$
CCD with 24 $\mu$m-pixels.
The chip has a read noise of 4.3 e$^-$ and was used with the standard
gain of 1.7 e$^-$/ADU. The peak system throughput for this setup is
roughly 5.5\%, estimated from standard-star observations (Bershady
\etal 2003b).

Given SparsePak's $\sim15$\arcsec\ center-to-center fiber spacing
and total area, we used 3 pointings along the galaxy's position angle 
to maximize spatial coverage and filling factor. 
Typical pointing offsets were $\sim 6$\arcsec.  
The observed galaxies have moderate sizes ($a \sim 2\farcm0$) and 
their velocity field can thus be mapped from center-to-edge. 
Total Sparsepak integrations consisted of 3 pointings $\times$ 2 900s 
exposures per pointing, for a total of 1.75 hours per galaxy. 
Multiple exposures at each position were used to identify and remove 
cosmic rays. 

Spectra obtained from SparsePak closely resemble WIYN Densepak or
Hydra spectra (\ie multi-fiber spectral data). Thus basic spectral 
extraction, flattening, wavelength calibration and sky
subtraction were done using the NOAO {\sl IRAF}\footnote{IRAF is 
 distributed by the National Optical Astronomy Observatories,
 which are operated by the Association of Universities for Research
 in Astronomy, Inc., under cooperative agreement with the National
 Science Foundation.} package {\it dohydra}. 
After basic reductions, we used a Gaussian line--fitting 
algorithm to measure Gaussian fluxes, widths, centers and centroid 
errors for H$\alpha$ emission lines (Andersen \etal 2003).
We rejected any line with a $S/N  < 5$.  More than 70\% of 
measured H$\alpha$ lines, 
even at the edge of the field, had significantly higher signal-to-noises,
with $S/N \gta 20$, yielding a mean centroiding error of only 
2.4 \kms\ for these 15 galaxies.

The V and I images were acquired at the WIYN 0.9-m telescope in 
f/13.5 mode (0\farcs43 pix$^{-1}$); integrations were 600s in each filter.
Isophotal brightness errors are $\lta 0.1$ \magarc at, or below,
26.5 \magarc in V and I.  
The imaging was obtained in non-photometric conditions (thin wisps
covered the Arizona desert sky) and thus cannot readily be merged 
into the Shellflow imaging data base.  Structural parameters can 
still be measured accurately, down to deep levels, as we discuss 
below.

Three previously observed SB (NGC 2540, UGC 5141, UGC 8229) and two 
SAB (NGC 3029, UGC 6895) Shellflow galaxies with available long-slit 
\ha spectra and V,I photometry were duplicated at the WIYN
telescopes for comparison.  These observations enable us to tie the
SparsePak velocity field information with the Shellflow long-slit
spectra obtained with the KPNO \& CTIO 4m telescopes + RC Spec 
(Courteau \etal\@2003b).

\subsection{Data Analysis}\label{sec:analysis}

Azimuthally-averaged surface brightness profiles were extracted for 
all the galaxies using ellipse fitting with a fixed center.  
To ensure a homogeneous computation of structural parameters and
color gradients, we use the position angles and ellipticities of 
our I-band isophotal maps to determine 
the SB profiles in the V-band.  The position angle and ellipticity 
are allowed to vary at each isophote.  Please refer to Courteau (1996) 
for details about our surface brightness extraction technique. 

Reduction techniques for the extraction of rotation curves 
from long-slit spectra are described in Courteau (1997).  
We shall simply state that the 1-D rotation curve is constructed 
by measuring an intensity-weighted centroid at each resolved 
major-axis \ha\ emission feature above a noise threshold. 
For the 2-D SparsePak data, a single, inclined, 
differentially rotating, circular disk model with a fixed center is 
used to fit the H$\alpha$ velocity fields (Andersen \& Bershady 2003).
Briefly, we assume a radially symmetric rotation curve and an axisymmetric 
velocity field. Using this smooth functional representation of the velocity 
field, we compared the model velocity field
to observations.  Parameters are varied using a multi-dimensional 
down-hill simplex method (Press \etal 1992) to minimize a $\chi^2$ 
statistic. Our velocity field model has nine free parameters: seven 
for the rotation
curve (see next paragraph), and two for inclination and position angle.
Two additional parameters account for positional offsets from differential
telescope pointing errors for each SparsePak position, yet in practice 
these parameters were consistent with zero and were thereafter not allowed 
to vary.

We parameterize the model used to fit the rotation curves of both
the 1-D (long slit) and 2-D (SparsePak) velocity field data with the 
following empirical function (Courteau 1997):
\begin{equation}
 v(r) =  v_0 + v_a \frac{(1+x)^{\beta}}{(1+x^\gamma)^{1/\gamma}},
\label{eq:vmod}
\end{equation}
where $x=1/R=r_t/(r-r_0)$, $v_0$ and $r_0$ are the velocity and spatial
centers of rotation, $v_a$ is an asymptotic velocity, and $r_t$ is a
transition radius between the rising and flat part of the rotation 
curve.  Solid-body rotation, or $v(r) \propto r$ (with $\partial 
v/\partial r \sim v_a/r_t$), is recovered for $|r-r_0| \ll r_t$, 
and flat rotation,
or $v(r) \propto v_a$, is achieved for $|r-r_0| \gg r_t$.
The term $\gamma$ governs the degree of sharpness of turnover, and $\beta$
can be used to model the drop-off or steady rise of the outer part of the 
rotation curve. 

Table~1 gives velocity field and structural parameters for the SparsePak data 
collected at WIYN in March 2002.  Listed are the number $N$ of velocity data
points, the kinematic and photometric inclinations, the kinematic and 
photometric position angles, the velocity fit parameters,  $v_a$, $r_t$, 
$\beta$, and $\gamma$ (see Eq.~1), the bar radius, $R_{bar}$ in the 
plane of the galaxy, the I-band scale length $h$ of the disk, and the 
recessional velocity of the galaxy, $v_0$.  The bar radius is defined as  
the location where the I-band surface brightness drops and/or position 
angle changes abruptly.  Disk scale lengths were determined as in 
MacArthur \etal (2003).
No photometric parameters are listed for IC 0784 which could not be observed 
at the telescope due to time and weather constraints. 

Appendix A contains rotation curves and extracted velocity fields 
(spider diagrams) for the WIYN/SparsePak galaxies.  The model rotation 
curves, based on Eq. (1), are a decent match to most extracted integral field 
velocity data points.  These models are shown mostly for illustrative 
purposes and for comparison with similar fits to rotation curves derived 
from long-slit spectra. They can also be used for future dynamical modeling. 

The overall impression from the comparison of velocity data for the 5 
Shellflow galaxies with long-slit 1-D and SparsePak 2-D rotation curves 
in Appendix A is very favorable.  For NGC 2540 (\Fig{n2540}), the 
1-D and 2-D velocity models are indistinguishable, owing in part to the 
very similar position angles and inclinations used to extract the 
velocity amplitudes.  The unbarred galaxy, NGC 3029 (\Fig{n3029}),
was re-observed for consistency check; again the velocity data and 
models agree very well within the measurement uncertainty.  NGC 5141 
shows only slight differences in the modeled RCs, and UGC 6895 and 
UGC 8229 show slightly larger differences in the inner slopes, perhaps 
caused by a misaligned slit.  
While the data distributions
agree within their respective scatter, the RC models predict different 
maximum rotation speeds, at the 10-20 \kms\ level. 
However, the basic impression to retain for this comparison is that 
long-slit and IFS rotation curves agree well within their measurement 
errors and intrinsic scatter and it can be assumed that line widths 
from 1-D rotation curves are a fair representation of the overall 
velocity field, even for barred galaxies. 
Close agreement between 1-D rotation curves from \ha\ long-slit spectra 
and major-axis rotation curves from Fabry-Perot (2-D) velocity fields 
was also demonstrated by Courteau (1997).

Another concern, when mapping the kinematic and dynamic structure 
of barred galaxies, is whether our diagnostics are affected by 
non-circular velocities, radial flows, and/or isophotal distortions.  
In order to assess the importance of non-circular motions, 
we have examined minor-axis rotation curves (not shown here for 
simplicity) and spider diagrams in Appendix A (see also Swaters
\etal 2003).  The minor-axis 
rotation curves are consistent with 10--20 \kms\ velocity dispersions 
of the turbulent gas with little hint of systematic deviations.  
The spider diagrams do show signs of non-circular motions, especially 
within $\sim 1.2R_{bar} (\simeq 1.5h_{disk}$).
However, beyond the extent or reach of the bar, most position-velocity
diagrams are symmetric about the major kinematic axis.  
With the exception of IC 2104 (\Fig{i2104}), a symmetric velocity 
pattern is recovered for all galaxies at, and beyond, $\r22$. 

The good match between 1-D and 2-D velocity fields and lack of 
significant non-circular motions at or beyond $\r22$ suggests that 
we can compare raw rotation speeds of barred and unbarred galaxies,
all other quantities being equal, without significant bias.  
This is what we do in \se{TFbar} for the Shellflow and SCII data. 
Any putative offset of the barred galaxies in the TF plane 
should not be due to systematic effects in the line widths.

Deprojection of velocity fields requires an inclination estimate.  
TF studies usually make use of photometric inclinations determined 
in the outer disk, away from a bar or spiral distortions, where 
ellipticities and position angles do not vary appreciably 
(\eg Courteau 1996, Beauvais \& Bothun 2001). 
We compare our SparsePak kinematic and I-band photometric inclination 
and position angle estimates in \Fig{inc_pa} and Table 1.  A position
angle offset would systematically lower the observed long-slit rotation,
and inclination differences could displace a galaxy in the TF plane. 
We find that galaxies with $i_{kin}>45$\deg\ show no appreciable 
inclination offset (within 3\deg\ rms) and a mild position angle 
offset (10\deg\ rms) between kinematic and photometric estimates. 
Position angle differences can be large for more face-on galaxies 
but our sample is too small to isolate systematic trends.

For galaxies with $i<35$\deg, photometric inclination angles are, on 
average, $\sim 12\%$ larger (more edge-on) than kinematic estimates.
Inclination offsets for the low-inclination unbarred galaxy NGC 3029
are large and can only be explained by model 
fitting (kinematic vs isophotal) differences, whereas excellent 
agreement is found for UGC 6895, a higher inclination ($i=45$\deg) 
unbarred galaxy. 

Note that our velocity model assumes circular, instead of elliptical,
orbits.  Kinematic inclinations are still precise enough to construct 
a TFR with small scatter ($\sigma_{\rm{TF}}\simeq0\magpoint3)$ even 
at very low inclinations (Andersen \& Bershady 2003).
It is however unclear which of the kinematic or photometric 
inclination is more ``representative'' of the disk projection on
the sky.
The inclination offset may result from a combination of kinematic 
modeling that favors more circular orbits and great sensitivity 
of the isophotal mapping technique to $m=2$ brightness perturbations.  
Spiral arms typically originate at the ends (ILR) of bars and retain 
a small pitch angle, highly noticeable in the brightness distribution,
hence the plausible bias towards higher photometric inclinations.  
These effects are especially acute when spiral arms are fully resolved. 

In a similar study, Sakai \etal (2000; \hub\ Key Project) find that
photometric and kinematic (radio synthesis mapping) inclination angles 
differ for barred galaxies.  Among the 21 calibrator galaxies in their 
TF sample, 7 are barred and their kinematic inclination angles are 
$\sim$ 10-15\% smaller than photometric inclinations.  Their barred 
galaxies all have $i_{phot} > 45$\deg.  However, inclination offsets 
for their unbarred galaxies are nearly absent.  Peletier \& Willner 
(1991) give radio and infrared inclination angles for 13 barred and 
unbarred nearby spirals with 27\deg$<i<$70\deg.  Radio synthesis 
inclinations are also $\sim$12\deg\ smaller than photometric estimates, 
but for all inclinations.  

To illustrate this potentially confusing situation, we plot 
in \Fig{bar_inc} the inclination difference, $\Delta i$ (kin -- phot),
against kinematic inclination for the galaxy samples considered above,
plus a sample of nearby, face-on, unbarred spiral galaxies
(Andersen 2001). 
At low inclinations, kinematic inclinations appear to be systematically 
lower (more face-on) than photometric inclinations, with a trend of 
increasing differences with decreasing inclination.  This is made very 
clear by examination of Andersen's data.  At high inclinations, both 
barred and unbarred galaxies have smaller inclinations offsets, 
apparently independent of inclination.  At these high inclinations,
the effect on the velocity deprojection is negligible ($< 5$\%).
It may be that SparsePak and photometric inclinations in 
these inclined galaxies are affected by extinction as higher 
opacity would naturally bias high optically-determined inclinations. 
However, the radio synthesis inclinations compiled in Sakai \etal are 
insensitive to dust and the inclination difference is most 
likely explained by modeling differences; 2-D velocity fields are 
modeled under the assumption of circular orbits and the larger 
kinematic inclinations at large inclination may result from an 
underestimate of the disk thickness.  
In general, with increasing inclination, photometric inclinations become
increasingly sensitive to the estimated disk thickness while velocity
fields (especially radio velocity fields) become increasingly affected
by warps and other non-circular motions. In any event, the inclination
differences at $i_{kin}>50$\deg\ are small ($<5$\deg) and do not affect 
our study.  Barnes \& Sellwood (2003) find a similar result for a 
sample of inclined galaxies with inferred photometric and kinematic 
(Fabry-Perot) inclinations. 

Opposite trends are found in the compilation of Peletier \& Willner (1991)
if all their data are considered.  Inclination offsets are large even at 
high inclinations.  This discrepancy however hinges on three galaxies,
NGC 4178, 4192, and 4216, that display various pathologies. NGC 4178 is
a very late type system, NGC 4192 has a strong warp in the outer disk, 
and NGC 4216 has a very pronounced dust-lane; these all make photometric 
measurements uncertain.  If we ignore the Peletier \& Willner data 
(bottom panel; \Fig{bar_inc}), we find that the transition threshold 
where kinematic inclinations becomes significantly lower than photometric
inclinations depends on type: $i_{kin} = 50$\deg, 40\deg, and 30\deg\ 
for barred, weakly-barred, and unbarred galaxies.

Clearly, a more extensive 2-D spectroscopic survey of barred and
unbarred galaxies in the near-infrared and radio will help address 
our general concerns about their dynamical structure and the limitations
of our modeling techniques.  Infrared imaging should also be secured
for extinction-free inclination measurements.  The measurement
of a ``true'' inclination of a galaxy is certainly ill-defined as it 
depends on the bandpass, dust extinction, the detector, reduction 
methods, and assumptions concerning the galaxy structure (e.g. presence 
of warps).  Yet, inclination angles from radio synthesis mapping may 
come closest to the most representative tilt angle of a galaxy on the sky. 

As we await more detailed comparisons of radio and optically 
determined inclinations, systematic differences between barred 
and unbarred galaxies can be avoided if we restrict our 
Shellflow and SCII samples to galaxies with $i_{phot}\gta50$\deg. 
Fortunately, all barred galaxies in our samples (Shellflow, SCII) already 
meet this criterion.  
We pursue our TF analysis with a discussion of Shellflow and SCII galaxies
below.  Our SparsePak sample will be reconsidered for TF analysis when 
calibrated imaging is available. 

\bigskip

\section{The Tully-Fisher Relation of Barred Galaxies}\label{sec:TFbar}

We use the ``Shellflow'' and ``SCII'' 
all-sky TF surveys to map the location of barred galaxies in the TF 
plane.  Shellflow includes 300 bright spiral (Sab-Scd) field galaxies in a 
shell bounded at $4500 < cz < 7000$ \kms, and SCII has 441 cluster 
spirals (Sa-Sd) spanning $5000 < cz < 19000$ \kms.

Shellflow galaxies were drawn from the
Optical Redshift Survey sample of Santiago \etal (1995) with inclinations
in the range [45\deg,78\deg], $m_{\rm{B}} \leq 14.5$, and
$|b| \geq 20^\circ$.  Interacting, disturbed, and some barred galaxies 
were rejected.  Rotation speeds from resolved \ha rotation curves
were measured at 2.2 disk scale lengths; the upper inclination limit
($i<78$\deg) reflects a desire to minimize
extinction effects in the inner parts of the rotation curve 
(\eg Courteau \& Faber 1988; Giovanelli \& Haynes 2002).
Deep I-band and V-band images were collected for each Shellflow
galaxy.  Disk scale lengths were obtained from B/D decompositions of
the azimuthally-averaged I-band surface brightness (SB) profile 
(Courteau \etal\@2003b).  

The SCII cluster galaxies were selected from CCD I-band images taken
at the KPNO and CTIO 0.9-m telescopes and classified by eye and by their 
bulge-to-disk ratio or concentration index. These galaxies have 
inclinations in the range
[32\deg,90\deg] and I-band magnitudes $12 \leq m_{\rm{I}} \leq 17$.
SCII line widths were measured from both \ha\ long-slit spectra
and \hi line profiles. 
SCII disk scale lengths were obtained by ``marking the disk'', or 
fitting the exponential part of the SB profile from 
${\sim}21$ I-\magarc to ${\sim}25$ I-\magarc (Dale \etal 1999).

Shellflow and SCII galaxies have $-20 \leq M^{\rm Shell}_{\rm{I}} \leq -24$ 
and $-18 \leq M^{\rm SCII}_{\rm{I}} \leq -24$, respectively.  Both TF 
calibrations are based on digital I-band imaging; $\VI$ colors, to test 
for $M/L$ variations 
and extinction effects, are available for the Shellflow sample only.
Deprojection of velocity widths uses photometric inclinations measured
in the outer disk where ellipticities and position angles do not vary
appreciably. 
Shellflow and SCII magnitudes are corrected for Galactic and internal
extinction and distances account for a Hubble expansion, bulk
flow model, and effects of incompleteness.  The exact choice
of distance scale does not affect our conclusions.

According to the RC3, 37\% of the Shellflow sample is barred
(SB types only).  In general, the proportion of galaxies with bars
of all sizes is even higher (Eskridge \etal\@2002)
but we are here only concerned with galaxies with the strongest bars;
\ie those with potentially the highest central baryon fraction.
Visual examination of the Shellflow galaxies
revealed only 6 strongly barred systems (at I-band); these
have $R_{bar}/h_{disk} \geq 1.2$, where $R_{bar}$ and $h_{disk}$ are the
size of the bar semi-major axis and disk scale length, respectively.
Visual examination of the SCII galaxies yielded 27 strongly barred galaxies
(D. Dale 2002; priv.\@ comm.)  In both samples, only barred
galaxies with $M_I \leq -20.4$ could be identified.
The Shellflow and SCII sub-samples of barred galaxies are by no means
complete, nor are the parent catalogs, and a significant number of 
bars will be missed especially at low magnitudes
and high inclinations where morphological identification 
becomes problematic.

Figs.~\ref{fig:ShellTFa} \& \ref{fig:SCIITF} show the 
distributions of rotational velocities and exponential 
scale lengths vs I-band absolute magnitudes for Shellflow 
and SCII galaxies.  Different symbols identify the full range of
spiral Hubble types; barred galaxies are further emphasized as
solid symbols with open circles.
Looking at the upper panel of \Fig{ShellTFa} for Shellflow galaxies,
one sees a small offset of barred galaxies from the mean TFR, 
consistent with these galaxies being systematically 
brighter for a given mass (line width).  The same statistically
loose trend for barred galaxies was observed by Sakai \etal (2000).
It could be explained if barred galaxies have higher star 
formation rates.  However, Phillips (1996) and Kennicutt (1999) 
find that global star formation rates in barred and unbarred 
galaxies of the same Hubble type are comparable.  
The TF offset, if real, might also be consistent with maximal 
disks being brighter than their dark-matter dominated counter-parts
at a given mass.

A clearer picture is obtained with the larger SCII sample (\Fig{SCIITF})
which shows no offset from the mean TFR for SCII barred galaxies. 
The combined velocity offset for the Shellflow and SCII barred galaxies 
in the two samples is $\langle \delta log V \rangle = -0.02 \pm 0.04$, 
consistent with no deviation of the mean TFR.  Note that photometric 
inclinations are used to deproject velocities in Shellflow
and SCII but using kinematic inclinations instead would simply imply
a readjustment of the TF zero-point.  Provided only one inclination
measure is used, the relative distribution of barred and 
unbarred TF galaxies is not affected by the precise choice of 
inclination (\se{analysis}).  Recall that all the Shellflow and 
SCII barred galaxies have $i>50$\deg\ and are not affected by a 
putative (kin-phot) inclination offset.  Furthermore, if we exclude 
the few unbarred galaxies that have $i<50$\deg\ from the Shellflow
and SCII samples, the TF distributions remain the same.  Thus, we 
conclude that {\it barred galaxies lie on the same TFR as 
unbarred galaxies.}  A similar realisation was also reached 
by Debattista \& Sellwood (2000). 

The kinship between barred and unbarred galaxies extends to other 
properties as well.  The lower panels of Figs.~\ref{fig:ShellTFa} 
\& \ref{fig:SCIITF} show no statistical differences in the scale 
lengths of barred and unbarred galaxies (for a given absolute magnitude). 
\Fig{ShellVMI} shows the color-magnitude diagram of Shellflow 
galaxies. Notwithstanding small statistics, barred and unbarred 
galaxies have similar colors, consistent with their having comparable 
star formation rates (Kennicutt 1999).  
MacArthur \etal (2003) find other similarities for structural parameters 
of barred and unbarred galaxies: their bar/bulge light profiles are 
close to exponential, and their ratio of bulge effective radius, $r_e$,
and disk exponential scale length, $h$, falls in the range 
$r_e/h=0.22 \pm 0.09$, expected for late Freeman Type I spirals. 

CR99 developed and applied a test for correlated scatter of the TFR.  
According to this test, pure stellar exponential (maximal) disks should 
deviate from the mean TF and luminosity-size (LS) relations in such a 
way that $\partialvr=-0.5$.  
Thus, strongly correlated TF/LS residuals for 
the barred spirals would support the suggestion that unbarred 
spirals have sub-maximal disks (high concentration halos) and 
that maximal disks are only found, on average, in barred spirals. 
A new analysis based on the Shellflow and 
SCII data sets yields residuals that are consistent with $\partialvr=0.0$
for both barred and unbarred galaxies.  CR99 found a similar result 
for the Courteau-Faber sample.  
This result further confirms earlier observations about spiral galaxies: 
barred and unbarred galaxies have similar physical properties and 
populate the same TF/LS relation and residual space.  It also shows 
that the TFR is fully independent of surface brightness (CR99),
a situation which may also result from the fine-tuning of virial 
parameters.  The analysis of the independence of surface brightness 
in the TFR, and a revised interpretation of the ``Courteau-Rix'' 
test in terms of virial parameter correlations, is presented in 
Courteau \etal\@(2003a).

\vspace{-.3cm}

\section{Discussion and Conclusion}\label{sec:conclusion}

We have tested the hypothesis that barred and unbarred spiral disks 
have different structural correlations, such as the Tully-Fisher relation,
with barred galaxies possibly having a higher luminous-to-dark matter 
fraction in their inner parts. 
New WIYN/SparsePak integral field spectroscopy and deep near-infrared 
photometry of barred and unbarred spirals allowed us to verify that 
non-circular motions are not significant at $\r22$ and that rotation 
curves from 1-D or 2-D spectroscopy are reliable beyond that radius.
Based on this result, and uniform inclination corrections for spiral 
galaxies with $i>50$\deg, we have compared the distribution of barred 
and unbarred galaxies in the TF plane from extensive redshift-distance 
surveys of galaxies and found no significant differences.  

For a given circular velocity, barred and unbarred galaxies have 
comparable luminosities, scale lengths, colors, and star formation 
rates\footnote{A comparative study Sheth \etal (2002) of the molecular 
gas properties of barred and unbarred galaxies in the BIMA Survey of 
Nearby Galaxies shows striking differences. However, their data (see 
their Fig.2) show less striking differences for the star formation rates 
between barred and unbarred galaxies but based on scanty information. 
More data are clearly needed to elucidate these questions!}.
This suggests that barred and unbarred galaxies are close members 
of the same family and do not originate from different evolutionary trees. 
Their structural duality may be understood if bars are generated by 
transient dynamical processes that are likely independent of the 
initial galaxy formation conditions. Their virial properties would 
otherwise be different. 

Very recent N-body simulations with the highest resolution have 
relaxed the notion that bars would grow in structures defined by
a narrow range of disk/halo parameters. Thus our comparisons
cannot be used to ascertain the notion that bars live mostly in spiral 
disks whose stellar fraction dominates the mass budget within the optical 
disk. Our results are however consistent with bright barred galaxies 
having similar dark matter fractions as do their unbarred cousins
(Debattista \& Sellwood 2000; Courteau \etal 2003a). 
Stellar velocity dispersions, which provide robust disk $M/L$ ratios,
hold the promise of breaking the disk/halo degeneracy in mass modeling
of barred and unbarred galaxies.

If the presence of bars in rotating disks is not directly related to 
their virial structure but rather to their local dynamical state, it
can surely be used as a signpost of galaxy evolution.
Given that bars may be just as important as mergers in shaping field 
disk galaxies, significant efforts should be invested in programs 
to probe differences between barred and unbarred galaxies. 
 Bars, which can be triggered spontaneously by the global dynamical 
instability of a rotationally supported disk, can also be induced by 
interactions with a satellite.  One might thus expect an {\it increase} 
of the fraction of barred disks at higher redshift, unless these younger 
disks are too dynamically hot to sustain bar unstable modes. 
Van den Bergh \etal (2002) studied the visibility of bars in the 
northern Hubble Deep Field (HDF-N) and reported a {\it dearth} of 
bars at $z>0.7$ in the rest-frame V-band.  Taken at face value, this 
could indicate a dependence of bar strength on the local galaxy 
density which grows with time.  However, a similar study
by Sheth \etal (2003) based on the NICMOS Deep Field reveals 
numerous strongly barred
galaxies up to $z=1.1$.  Extinction effects in the bluer band explored 
by van den Bergh \etal (2002) thus thwarted their ability 
to detect dust enshrouded bars.  
Given the detection of stable disks beyond $z\sim1.3$  (van Dokkum 
\& Stanford 2001; Genzel \etal 2003), it is thus reasonable to posit
the existence of bars at comparably high redshifts. 
The cosmological volumes sampled in two HDF studies above are very small 
and robust statistics on the barredness of galaxies with look-back 
time awaits wider coverage and more extensive sky surveys, especially
with telescopes like ALMA within the next decade.  

Closer to home and on shorter time scales, our comparison of a few 
dozen barred galaxies with TF samples of unbarred disks should soon
be superceded, it is wished, by systematic studies of structural and 
environmental properties of thousands of barred and unbarred galaxies 
in the SLOAN and 2MASS galaxy catalogs.  Only with such large-scale,
systematic local investigations can we make significant progress in 
mapping galaxy evolution at high-redshift and linking the near and 
far-field Universe. 

\vspace{-0.6cm}
\acknowledgments

We are grateful to Daniel Dale for information about the SCII data base
and to Anatoly Klypin, Jerry Sellwood and Ben Weiner for comments about 
N-body simulations of barred galaxies.  Constructive suggestions by the
referee improved the flow of the paper. 
This research has made use of the NASA/IPAC extragalactic
database (NED) which is operated by the Jet Propulsion Laboratory,
California Institute of Technology, under contract with the National
Aeronautics and Space Administration.  SC wishes to acknowledge his
colleagues on the Shellflow team (Marc Postman, David Schlegel, and 
Michael Strauss) for permission to use previously unpublished results.
SC and LAM acknowledge financial
support from the National Science and Engineering Council of Canada.
MAB acknowledges financial support from NSF grant AST-9970780.
SC would also like to thank the Max-Planck Institut f{\" u}r Astronomy
in Heidelberg and the Max-Planck Institut f{\" u}r Astrophysik in Munich
for their hospitality while much of this paper was cooked up. 

\newpage

\appendix

\section{Rotation Curves, Velocity Fields and I-band images for the WIYN02 
  Sample}\label{appendixA}

This section shows long-slit (1-D) and SparsePak (2-D) velocity fields 
for all galaxies observed at WIYN in March 2002.  See \se{WIYN} for 
details about the sample and data analysis.  Shown for each galaxy 
are the position-velocity contours (``spider'' diagrams) in
the upper window, superimposed on the galaxy I-band image, and 
in the lower window, the SparsePak velocities (in the plane of the sky,
\ie not corrected for projection effects).  The velocity data were 
extracted according to various techniques described in the text and, 
whenever available, matching rotation curves are shown from the 
Shellflow collection of long-slit spectra.  

Smoothed versions of the observed velocity field were 
produced using the {\sl patch} routine within the GIPSY analysis package
(van der Hulst \etal 1992; Vogelaar \& Terlouw 2001).
The SparsePak velocity field shown in the lower window is extracted 
from a model which includes inclination,
PA, disk center, rotation velocity, scale length, and systemic velocity. 
The parameterization of the velocity field is given by \Equ{vmod}.  

SparsePak \ha\ position-velocity diagrams are constructed using
2 representations of the 2-D velocity field: The first includes all 
measurements with a simulated 6\arcsec\ ``slit'' for the
best-fit kinematic position angle (filled triangles). The second SparsePak
rotation curve uses all measured velocities within $\pm60$ \deg\ of the
kinematic major axis {\it in the inclined plane} of the galaxy (open squares).
Using the modeled kinematic inclination and position angle, we can project
each measured rotation velocity onto the major axis. This second, "wedge,"
approach is relatively insensitive to inclination-induced beam smearing
which affects the simulated slit measurements. However, the wedge does
not spatially sample the inner 10\arcsec\ as well as the slit. Our
best fit model (solid line) is adjusted for beam smearing induced by
the $\sim 5$\arcsec\ fibers of SparsePak.  When comparing this model to
the data, remember that the simulated slit data (filled triangles) have
not been projected onto the major axis; the magnitude of these velocities
serve only as a lower limit. Thus, a black triangle in the center of the RC 
that does not have a corresponding open box at the same radius implies 
that the center of that fiber lies more than 60\deg\ from the major axis
and its azimuth correction is large (greater than 2). 
The velocity models
based on \Equ{vmod} trace the open boxes only. 
Further details about velocity field modeling are given in \se{analysis}.

\clearpage

\begin{figure}[!t]
\vskip 15.0cm
{\includegraphics{Table1.ps}}
\end{figure}
\clearpage

\begin{figure}
\plotone{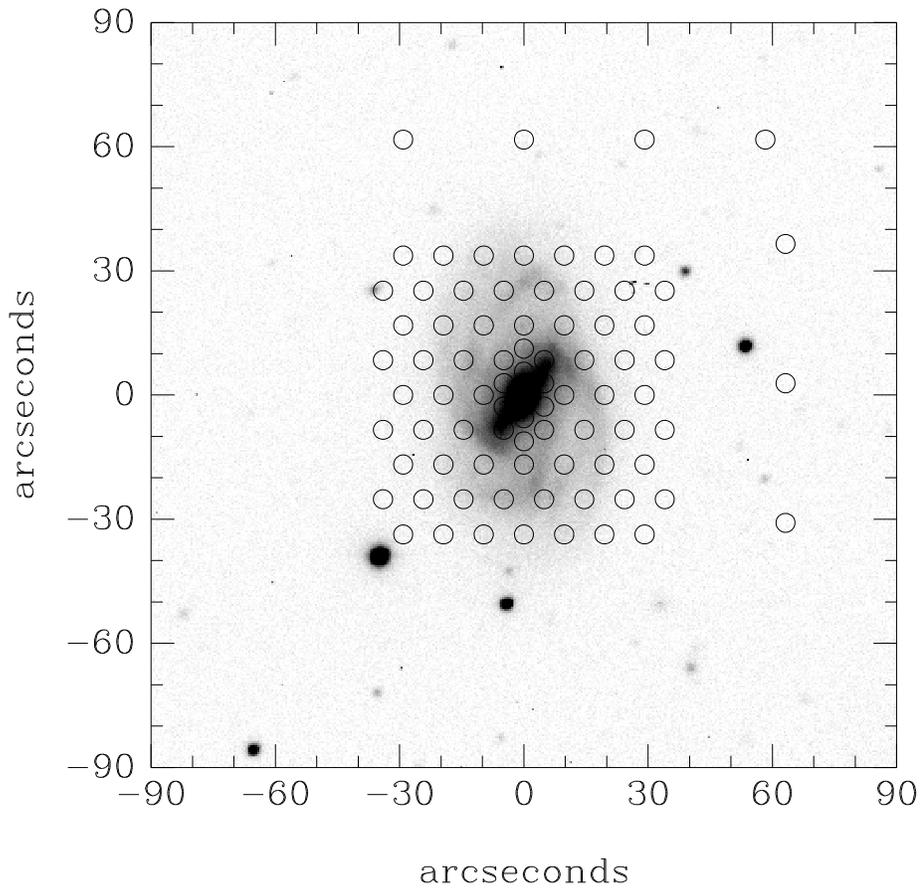}
\caption{SparsePak fiber footprint for one pointing overlayed on our 
         CCD I-band image for the SBbc galaxy UGC 5141. 
\label{fig:fibers}}
\end{figure}
\clearpage


\begin{figure}
\plotone{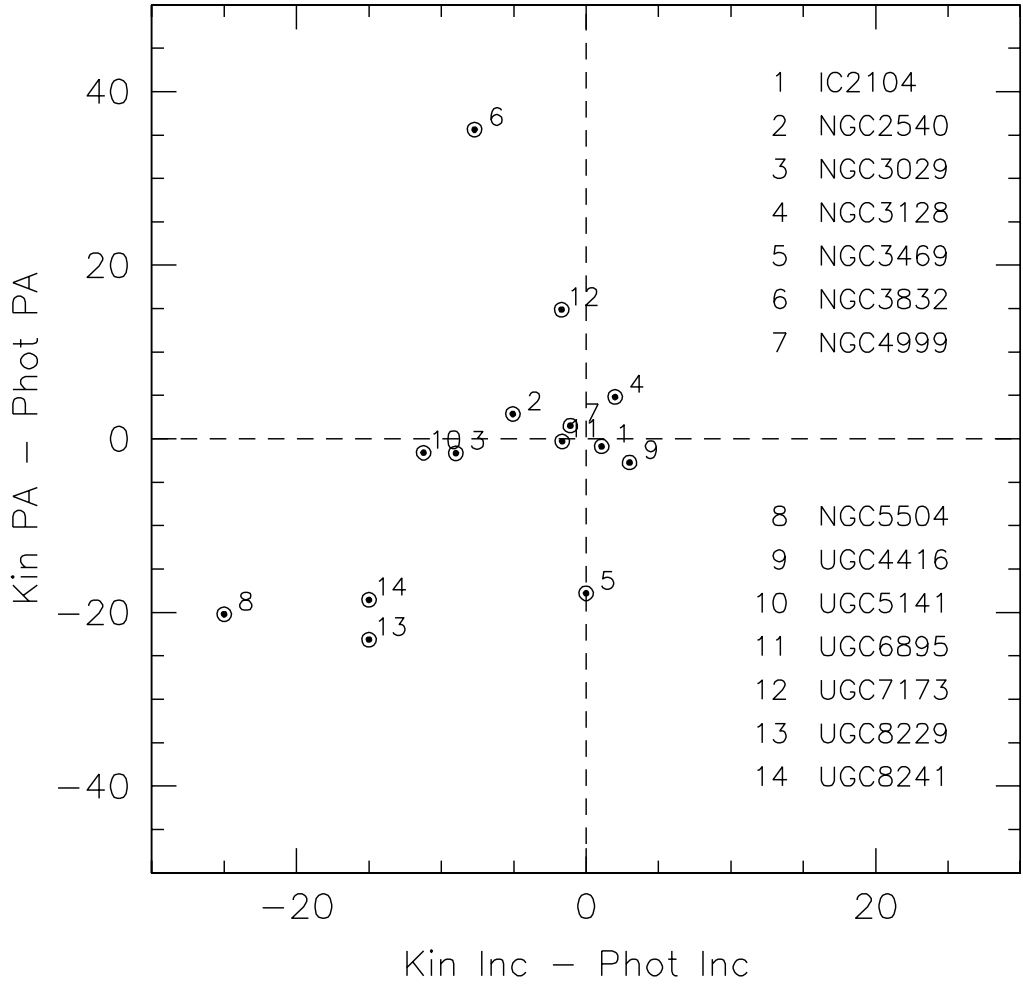}
\caption{Differences in measurements of kinematic and photometric  
 position angles and inclinations for galaxies with available 
 2-D velocity fields and I-band imaging.  Inclinations shown next
 to the galaxy names correspond to the kinematic and photometric 
 estimates, respectively.  Inclination differences are larger for
 progressively face-on orientations.  
\label{fig:inc_pa}}
\end{figure}
\clearpage

\begin{figure}
\plotone{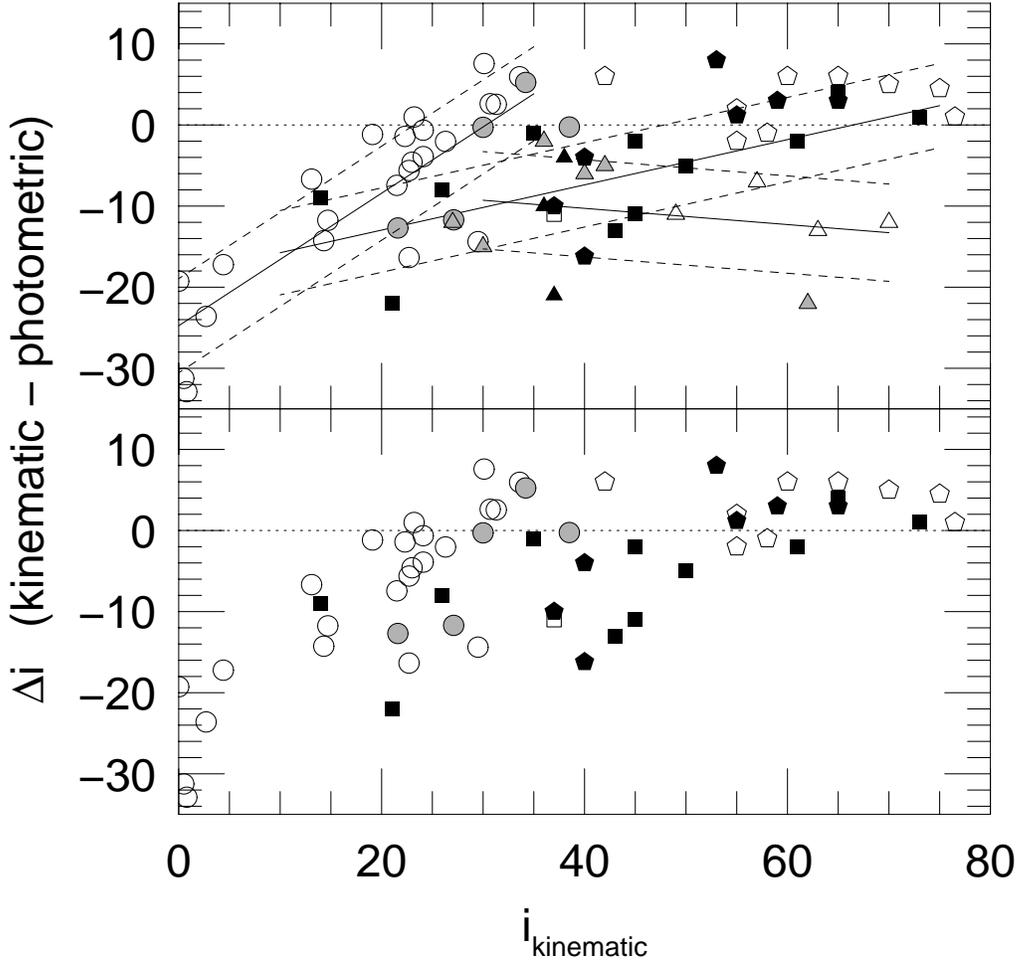}
\vspace{-4cm}
\caption{Difference between kinematic and photometric
 inclinations vs kinematic inclination for four galaxy 
 samples.  The point types are squares (this study); 
 circles (Andersen 2001); triangles (Peletier \& Willner 1991); 
 and pentagons (Sakai \etal 2000).  
 Open, gray-filled, and black-filled symbols represent 
 unbarred, weakly-barred, and strongly barred galaxies, 
 respectively.  The top panel shows simple regressions 
 to the Courteau \etal (this study), Andersen, and 
 Peletier \& Willner samples (independently).  We exclude 
 the Peletier \& Willner sample in the bottom sample. 
\label{fig:bar_inc}}
\end{figure}
\clearpage

\begin{figure}
\plotone{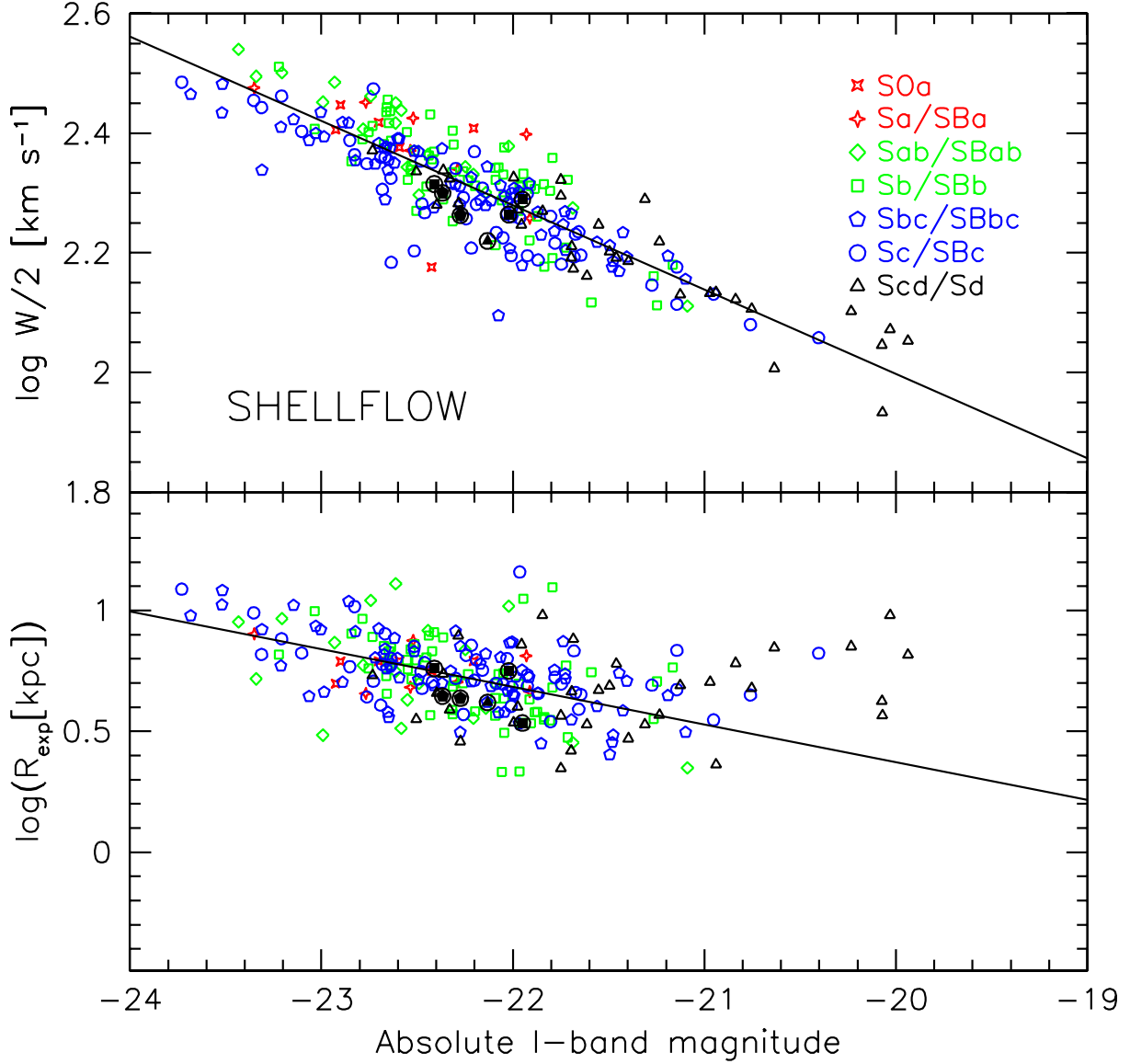}
\caption{Line width-luminosity (top) and size-luminosity (bottom) diagrams
 for Shellflow galaxies.  Line widths are measured at 2.2 disk scale lengths
 and disk scale lengths are obtained from B/D decompositions of the surface 
 brightness profile.  Barred galaxies have filled symbols consistent with 
 their Hubble type and are further emphasized with an open circle.  
 Barred galaxies lie below the mean TFR, appearing to be
 systematically brighter for their rotational velocity.  As in
 Sakai \etal (2000), this is a small number artifact. 
 The solid line is a fit from our data$-$model minimization 
 technique (Courteau \etal 2003a). 
\label{fig:ShellTFa}}
\end{figure}
\clearpage

\begin{figure}
\plotone{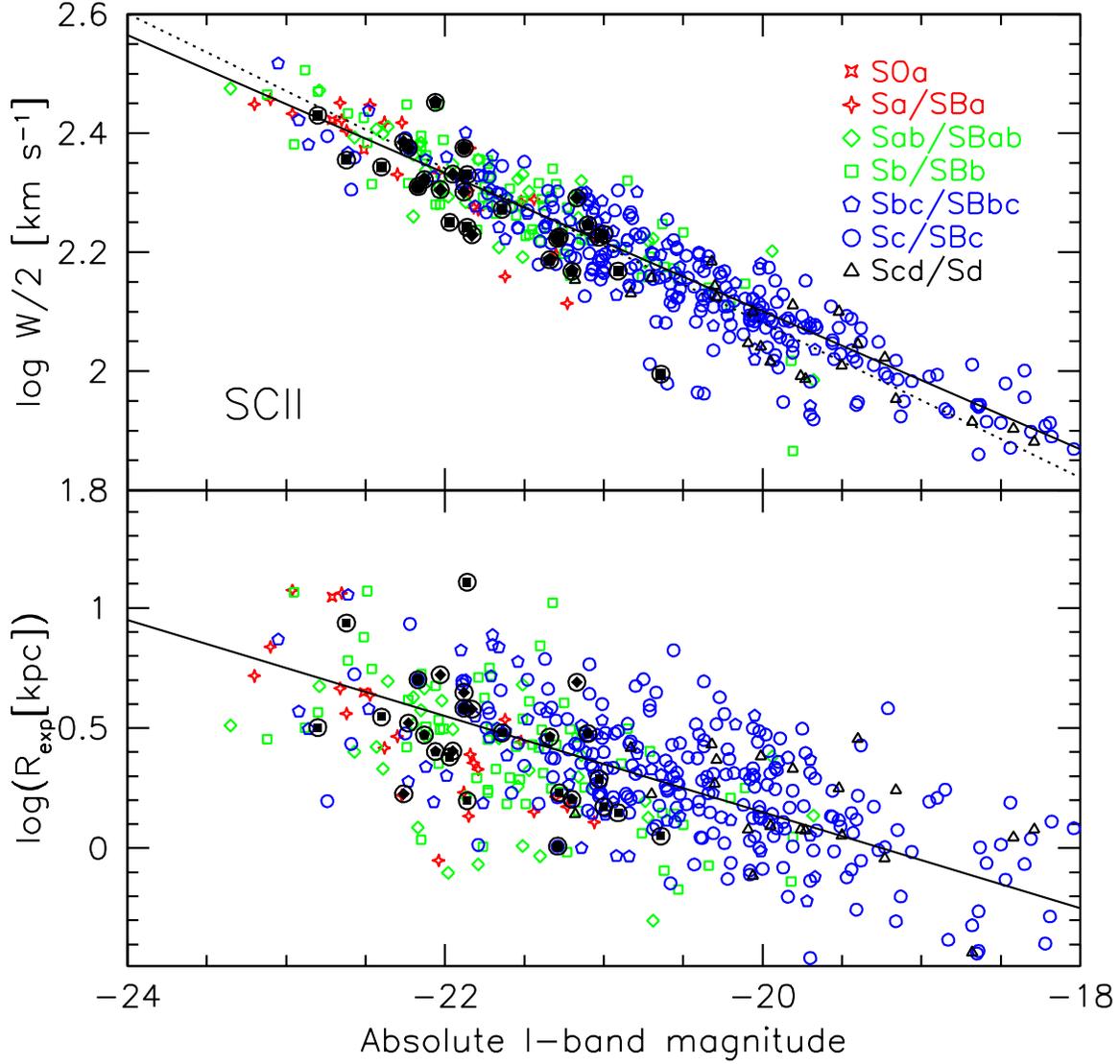}
\caption{Line width-luminosity (top) and size-luminosity (bottom) diagrams
 for SCII galaxies. Line widths are measured from \ha rotation curves and 
 \hi line widths and disk scale lengths are measured using the ``marking 
 the disk'' technique (see text).  Symbols are as in \Fig{ShellTFa}.  
 The TFR is the same for barred and unbarred galaxies.  The solid and 
 dashed lines show data$-$model minimization fits from 
 Courteau \etal (2003a) and Dale \etal (1999), respectively. 
\label{fig:SCIITF}}
\end{figure}
\clearpage

\begin{figure}
\plotone{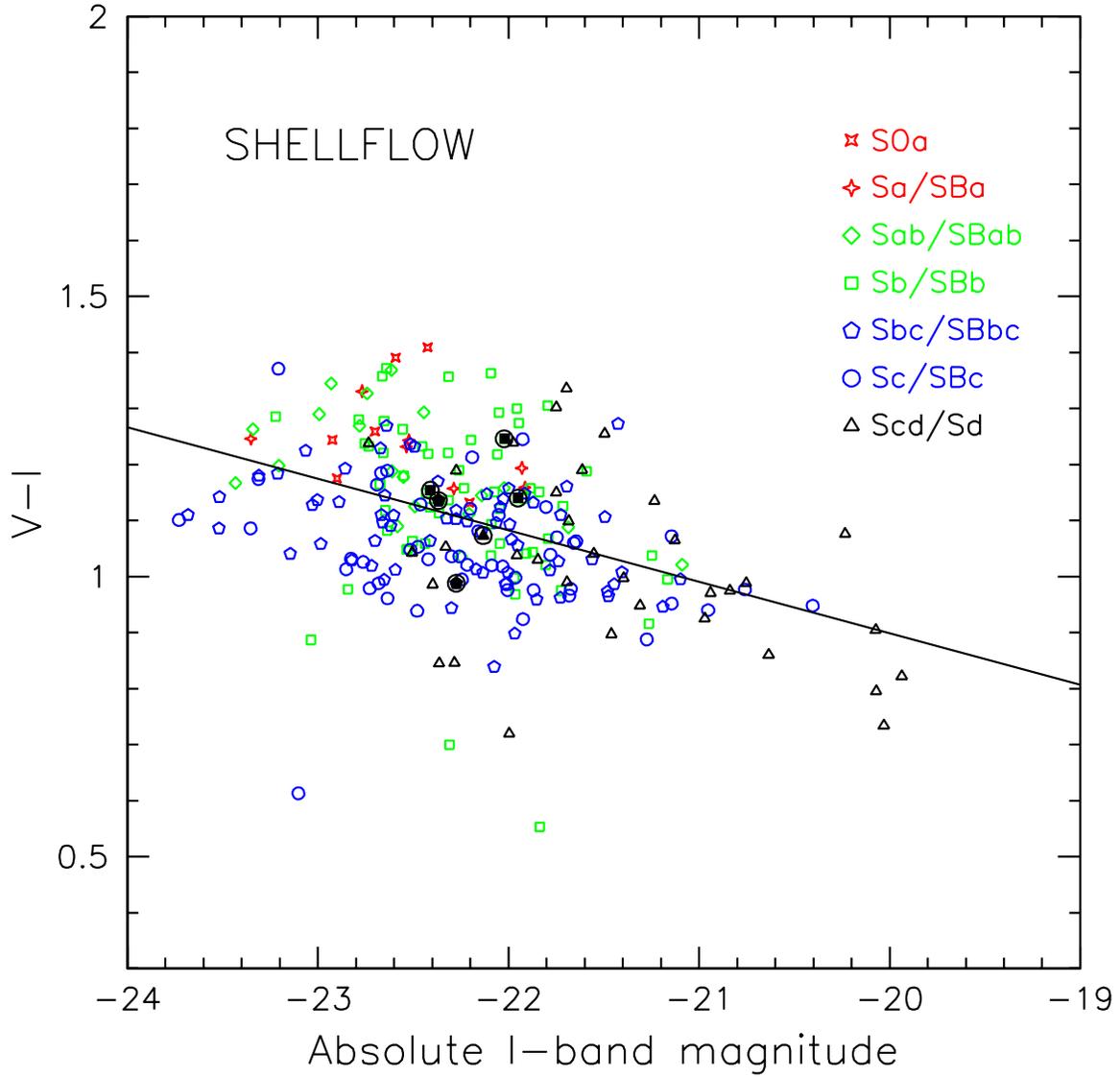}
\caption{Color-magnitude diagram for Shellflow galaxies.
 Barred galaxies have mean colors consistent with the general 
 spiral population.  Symbols are as in \Fig{ShellTFa}.
\label{fig:ShellVMI}}
\end{figure}
\clearpage 



\begin{figure}[]
 \vskip 7in
 {\includegraphics{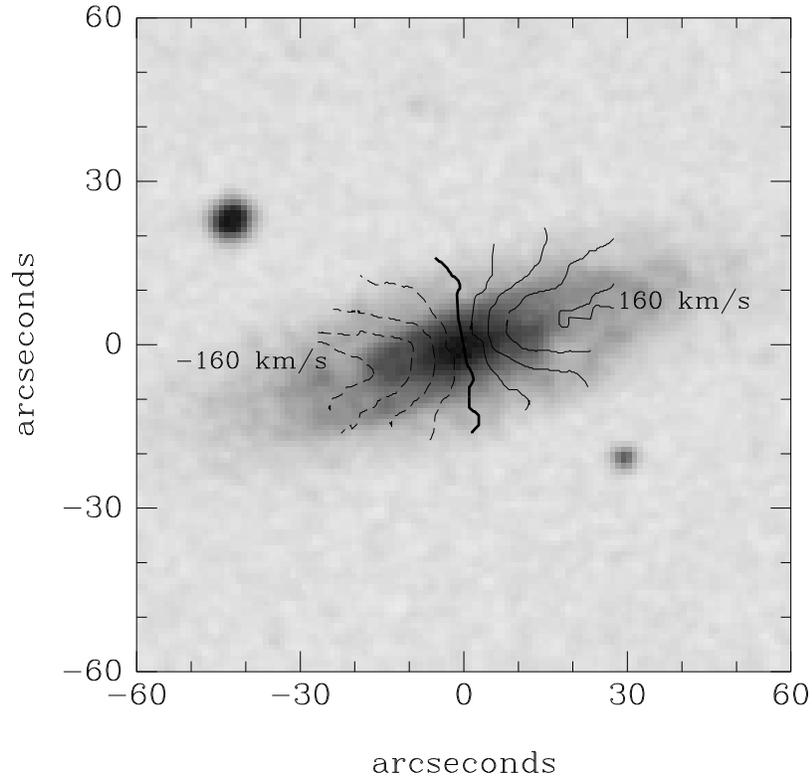}}
 {\includegraphics{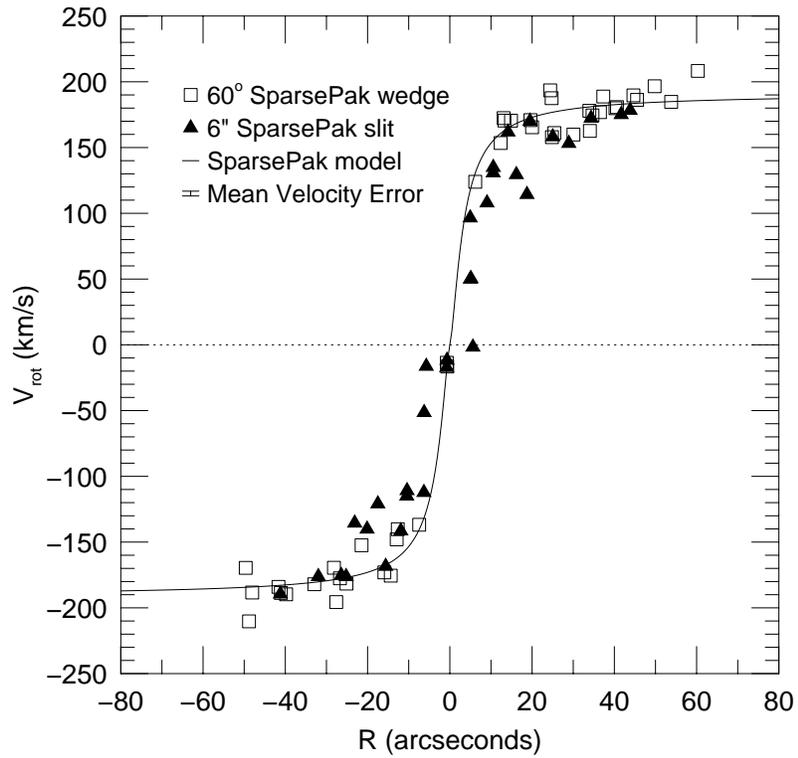}}
 \vskip 4cm
\caption{Velocity contours and I-band image (top)
 and rotation curve data with velocity model (bottom) 
 for IC 784.
\label{fig:i0784}}
\end{figure}

\begin{figure}[]
\caption{Velocity contours and I-band image (top)
 and rotation curve data with velocity model (bottom) 
 for IC 2104.  This galaxy has a pathological velocity 
 field with significant non-circular motions, a continuously
 rising rotation curve, and a small inner velocity bump 
 representative of a strong bar and/or bulge (that is poorly
 matched by the velocity model).
\label{fig:i2104}}
\end{figure}

\begin{figure}[]
\caption{Velocity contours and I-band image (top)
 and rotation curve data with velocity model (bottom) 
 for NGC 2540.
\label{fig:n2540}}
\end{figure}

\begin{figure}[]
\caption{Velocity contours and I-band image (top)
 and rotation curve data with velocity model (bottom) 
 for the unbarred galaxy NGC 3029.
\label{fig:n3029}}
\end{figure}

\begin{figure}[]
\caption{Velocity contours and I-band image (top)
 and rotation curve data with velocity model (bottom) 
 for NGC 3128.
\label{fig:n3128}}
\end{figure}

\begin{figure}[]
\caption{Velocity contours and I-band image (top)
 and rotation curve data with velocity model (bottom) 
 for NGC 3469.
\label{fig:n3469}}
\end{figure}

\begin{figure}[]
\caption{Velocity contours and I-band image (top)
 and rotation curve data with velocity model (bottom) 
 for NGC 3832.
\label{fig:n3832}}
\end{figure}

\begin{figure}[]
\caption{Velocity contours and I-band image (top)
 and rotation curve data with velocity model (bottom) 
 for NGC 4999.
\label{fig:n4999}}
\end{figure}

\begin{figure}[]
\caption{Velocity contours and V-band image (top)
 and rotation curve data with velocity model (bottom) 
 for NGC 5504.  The I-band image was not available. 
\label{fig:n5504}}
\end{figure}

\begin{figure}[]
\caption{Velocity contours and I-band image (top)
 and rotation curve data with velocity model (bottom) 
 for UGC 4416.  The vertical trace in the upper image 
 is due to an internal image reflection. 
\label{fig:u4416}}
\end{figure}

\begin{figure}[]
\caption{Velocity contours and I-band image (top)
 and rotation curve data with velocity model (bottom) 
 for UGC 5141.
\label{fig:u5141}}
\end{figure}

\begin{figure}[]
\caption{Velocity contours and I-band image (top)
 and rotation curve data with velocity model (bottom) 
 for the weakly barred galaxy UGC 6895.
\label{fig:u6895}}
\end{figure}

\begin{figure}[]
\caption{Velocity contours and I-band image (top)
 and rotation curve data with velocity model (bottom) 
 for UGC 7173.
\label{fig:u7173}}
\end{figure}

\begin{figure}[]
\caption{Velocity contours and I-band image (top)
 and rotation curve data with velocity model (bottom) 
 for UGC 8229.
\label{fig:u8229}}
\end{figure}

\begin{figure}[t]
\caption{Velocity contours and I-band image (top)
 and rotation curve data with velocity model (bottom) 
 for UGC 8241.
\label{fig:u8241}}
\end{figure}

\end{document}